\documentstyle[prc,aps,epsfig,multicol]{revtex}
\draft
\tightenlines
\begin{document}
\title{$\Xi$ production at AGS energies}
\author{Subrata Pal and C. M. Ko}
\address{Cyclotron Institute and Physics Department,
Texas A\&M University, College Station, Texas 77843-3366}
\author{J. M. Alexander, P. Chung, and R. A. Lacey}
\address{Department of Chemistry, State University
of New York at Stony Brook, Stony Brook, New York 11794-3400}

\maketitle

\begin{abstract}
A relativistic transport model is used to study $\Xi^-$ production
in $2-11A$ GeV Au+Au collisions. Introducing the strangeness-exchange 
reactions between antikaons and hyperons as the sources for $\Xi^-$,
we find that the cascade yield in these collisions is in reasonable 
agreement with the data. Although the $\Xi^-$ abundance does not 
reach chemical equilibrium unless the cross section for 
strangeness-exchange reactions is enhanced by six times, it 
exhibits the strongest enhancement with increasing centrality of 
collision and with increasing beam energy.
\medskip

\end{abstract}

\begin{multicols}{2}
 
\section{introduction}

One of the most challenging problems in relativistic heavy-ion
collisions is to explain the observed enhancement of singly strange and 
multistrange hadrons. The shorter equilibration times for strange quarks
in a quark-gluon plasma (QGP) than the strange hadrons in hadronic matter 
had led to the conjecture \cite{rafelski} that enhanced production of 
(multi)strange hadrons might be an indication of QGP formation in the 
early stage of the collision. However, a firm conclusion of strangeness
enhancement as a signal for the QGP can only be established 
if baseline hadronic processes fail to explain the strange particle 
abundances. For the singly-strange kaons and lambdas, it was found
that the observed enhancement at both the AGS \cite{abbot} and the 
SPS \cite{bear} energies could indeed be understood within a hadronic 
scenario \cite{transport}. This scenario, on the other hand, cannot 
explain the enhanced production of the multistrange $\Xi$ and $\Omega$ 
(and their antiparticles) observed in Pb+Pb collisions at SPS energies 
\cite{soff,pal1}. The latter enhancement has thus been associated 
with production mechanisms which include possible QGP formation\cite{rafel}.
The doubly strange $\Xi$ has also been measured in heavy ion
collisions at the AGS \cite{paul}.  Compared to that expected 
from primary nucleon-nucleon collisions, the observed enhancement is 
even more pronounced than in the SPS experiments. Although this is
not surprising as experimental data have shown that the ratio of 
strange to nonstrange particles (such as $K^+/\pi^+$) in A+A relative 
to p+p collisions increases gradually from RHIC energy to lower 
beam energies at SIS \cite{dunlop}, it is interesting to find out 
if the measured $\Xi$ yield in this collision is consistent with 
the hadronic or the QGP scenario. In this paper, we shall study
the production of multistrange $\Xi^-$ from Au+Au collisions at the AGS
energy of $6A$ GeV using the relativistic hadronic transport model
ART \cite{art}. We shall show that strangeness-exchange reactions between
the antikaon and lambda(sigma) in hadronic matter alone can explain the 
measured $\Xi$ yield \cite{paul}, and exotic processes for 
multistrange particle production at the AGS energies are thus not required.
 
\section{The ART model and the strangeness-exchange reactions}

The ART model is a hadronic transport model that includes 
baryons such as $N$, $\Delta(1232)$, $N^{*}(1440)$, $N^{*}(1535)$, 
$\Lambda$, $\Sigma$, and mesons such as $\pi,~\rho,~\omega,~\eta,~K,~K^*$. 
Both elastic and inelastic collisions among most of these particles 
are included by using the experimental data from hadron-hadron collisions. 
The ART model has been quite successful in explaining many experimental
observations, including the surprisingly large kaon antiflow
\cite{chung,pal} in heavy ion collisions at AGS energies. The ART model 
also allows us to understand whether or not the strongly interacting matter
formed in these collisions reaches chemical and/or thermal equilibrium.
In the present study, we extend the ART model to include
perturbatively the $\Xi$ particle as in the studies for other rare
particles using the transport model \cite{pal1,randrup,fang}.

In hadronic matter, the $\Xi$ particle is mainly produced via the 
strangeness-exchange reactions $\bar K(\Lambda,\Sigma)\to\pi\Xi$ 
because the contribution from the associated production
reactions $NN\to\Xi NKK$ and $\pi (\Lambda,\Sigma)\to K\Xi$
are Okubo-Zweig-Iizuka suppressed. Since there is no empirical 
information on the strangeness-exchange reaction for $\Xi$ production, 
we use the cross section obtained from the gauged flavor
$SU(3)$-invariant Lagrangian in the coupled-channel approach \cite{chli}. 
Because the large threshold for final states with the $\eta$ particle, 
only the reactions $\bar K(\Lambda,\Sigma)\to\pi\Xi$ are considered
here. Writing the spin- and isospin-averaged cross sections for these 
reactions as
\begin{eqnarray}
\sigma _{\bar{K}\Lambda \rightarrow \pi \Xi }&=&\frac{1}{4}
\frac{p_{\pi }}{p_{\bar{K}}}|M_{\bar{K}\Lambda \rightarrow \pi \Xi}|^{2}, 
\nonumber\\
\sigma _{\bar{K}\Sigma \rightarrow \pi \Xi }&=&\frac{1}{12}
\frac{p_{\pi }}{p_{\bar{K}}}|M_{\bar{K}\Sigma \rightarrow \pi \Xi }|^{2},
\qquad
\end{eqnarray}
where $p_{\bar K}$ and $p_\pi$ are initial antikaon and final pion
momenta in the center-of-mass system, the theoretical cross sections
of Ref.\cite{chli} are found to be well fitted by the following squared 
invariant matrix elements $|M_{\bar K\Lambda\to\pi\Xi}|^{2}$ and 
$|M_{\bar K\Sigma\to\pi\Xi}|^{2}$ \cite{chen}:
\begin{eqnarray}
|M_{\bar{K}\Lambda \rightarrow \pi \Xi }|^{2}&=&34.7\frac{s_0}{s}
~{\rm mb},\nonumber\\
|M_{\bar{K}\Sigma \rightarrow \pi \Xi|}|^{2}&=&318
\left(1-\frac{s_0}{s}\right)^{0.6}\left(\frac{s_0}{s}\right)^{1.7}~{\rm mb}.
\label{param}
\end{eqnarray}
In the above, the threshold energy $s_0^{1/2}$ in the center-of mass system
is 1.611 GeV and 1.688 GeV for the reactions $\bar K\Lambda\to\pi\Xi$
and $\bar K\Sigma\to\pi\Xi$, respectively. We note that except near 
threshold, these cross sections are of the order of $5$-$10$ \textrm{mb}.

The cross sections for the inverse reactions $\pi\Xi\to\bar K\Lambda$
and $\pi\Xi\to\bar K\Sigma$, which are needed for treating $\Xi$ 
annihilation, are related to those for $\Xi$ production by the
principle of detailed balance, i.e., 
\begin{eqnarray}
\sigma_{\pi\Xi\rightarrow\bar{K}\Lambda}&=&\frac{1}{3}
\frac{p_{\bar{K}}^2}{p_{\pi}^2}\sigma_{\bar K\Lambda\to\pi\Xi},\nonumber\\
\sigma_{\pi\Xi\rightarrow\bar{K}\Sigma}&=&\frac{p_{\bar{K}}^2}{p_{\pi}^2}
\sigma_{\bar K\Sigma\to\pi\Xi}.
\end{eqnarray}

We have also included the production of the resonance $\Xi^*(1530)$ 
via the reaction $\pi\Xi\to \Xi^*(1530)$, and its cross section is 
given by a Breit-Wigner form with the empirical width of 
$\Gamma_{\Xi^*}=9.5$ MeV. The cross section for the inverse reaction
of $\Xi^*(1530)$ decay, i.e., $\Xi^*(1530)\to\pi\Xi$, is also obtained 
from the detailed balance relation. 

\section{results}

\subsection{Time evolution of hadron abundances}

\begin{figure}[ht]
\centerline{\epsfig{file=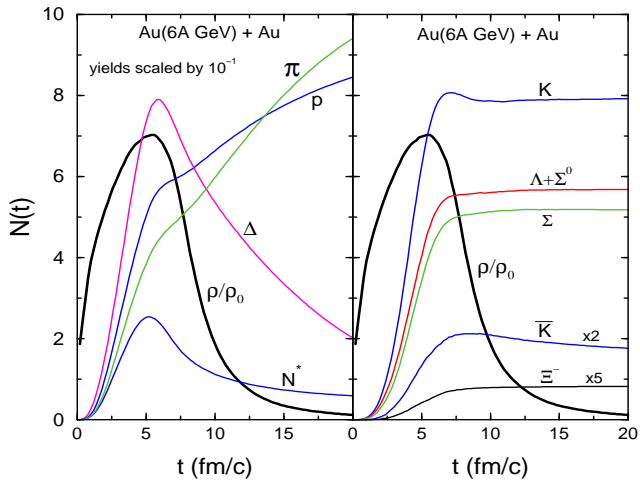,width=3.3in,height=2.5in,angle=0}}
\caption{Time evolution of central density and hadrons at midrapidity $|y|<0.5$
in $6A$ GeV Au+Au collisions at impact parameter $b\leq 3$ fm in the
ART model.}
\label{tevol}
\end{figure}

Figure \ref{tevol} displays the time evolution of midrapidity hadrons 
and the central density obtained in the ART model for central Au+Au 
collisions at $6A$ GeV. The abundances for baryon resonances (left panel) 
attain peak values at the maximum compression stage of collision at a
time $t \approx 6$ fm/c after the initial contact of the colliding nuclei. 
During subsequent decompression, the resonances 
decay to produce the stable nucleons and pions. The kaons produced
from baryon-baryon and pion-baryon collisions in ART model,
tends to saturate soon after the maximum compression stage (right panel).
On the other hand, the antikaon yield steadily decreases during the expansion 
stage due to their strong absorption with the abundant nucleons at the AGS 
energies via $\bar K N \to \pi (\Lambda,\Sigma)$. The hyperons 
($\Lambda, \Sigma, \Xi$ plus their respective resonances) are produced within 
8 fm/c when the density is high. Thereafter, the abundances are rather 
insensitive to hadronic scatterings, if any. We find that the
strangeness-exchange reactions between antikaons and hyperons lead 
to an appreciable production of $\Xi$.

\begin{figure}[ht]
\centerline{\epsfig{file=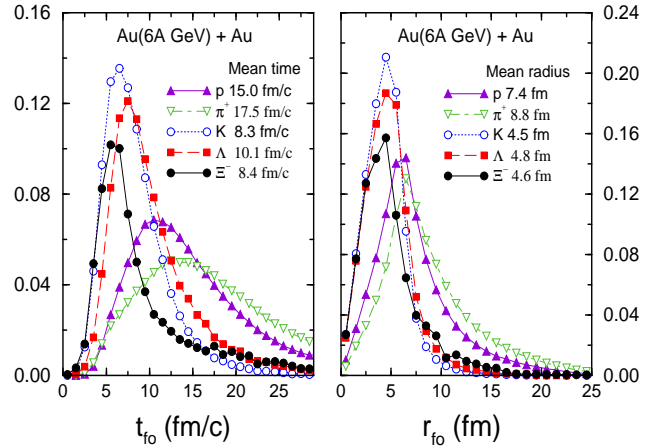,width=3.3in,height=2.5in,angle=0}}
\caption{Time and transverse radius distributions of midrapidity ($|y|<0.5$) 
hadrons at freeze-out in $6A$ GeV Au+Au collisions at impact parameter 
$b\leq 3$ fm from ART model.}
\label{frzout}
\end{figure}

The time and spatial distribution of midrapidity particles at freeze-out are 
shown in Fig. \ref{frzout}. Most of the $\Xi$'s freeze-out between 
3 and 7 fm/c, whereas the freeze-out distribution of protons and pions 
is centered at about 12 fm/c (left panel). Due to the fact that strange 
hadrons have much smaller scattering cross sections with nucleons 
and pions, they decouple quite early from the system. The dominantly 
soft collisions among the nonstrange hadrons lead to a long tail in 
the nucleon and pion distributions.  The transverse radius
distribution (right panel) indicates that the $\Xi$ and other hyperons 
have source sizes similar to the initial source.

\vspace{-0.6cm}

\subsection{Impact parameter dependence}

\begin{figure}[ht]
\centerline{\epsfig{file=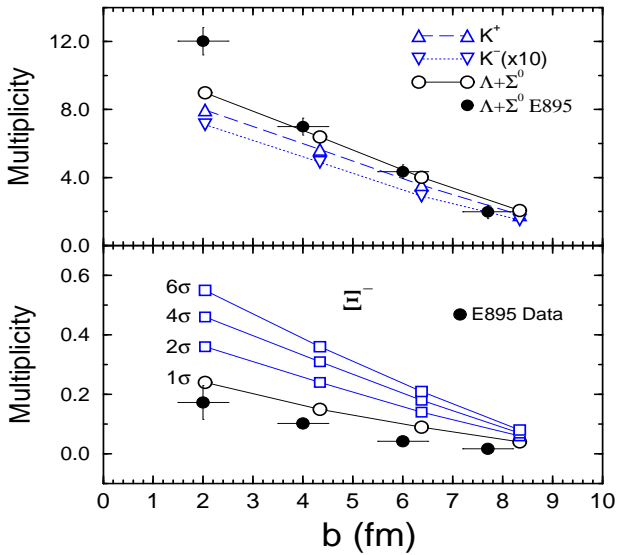,width=3.2in,height=3.0in,angle=0}}
\vspace{-0.1cm}
\caption{Calculated impact parameter dependence for the total yield of
$K^+$, $K^-$, $\Lambda+\Sigma^0$ (top panel) and $\Xi^-$ (bottom panel) 
in Au+Au collisions at $6A$ GeV. The dependence of $\Xi^-$ production 
on the cross section $\sigma$ for $\bar K (\Lambda,\Sigma) \to\pi\Xi$ 
is also shown. The E895 data for the $\Lambda+\Sigma^0$ (top panel) 
and $\Xi$ yield (bottom panel) are shown by solid circles.}
\label{bmpact}
\end{figure}

In Fig. \ref{bmpact}, we show the impact parameter dependence of the total
yield of different particle species obtained from the ART model for
Au+Au collisions at $6A$ GeV. Except for very peripheral collisions,
a $K^-/K^+$ ratio of 0.09 is obtained and found to be independent of
collision centrality. This trend is also observed experimentally
from  $\sim 11A$ GeV up to RHIC energies \cite{ahle,harris}.
Within a statistical approach, this implies that $K^+$ and $K^-$
have similar freeze-out volumes. Indeed, in the transport calculation
kaons and antikaons exhibit similar mean freeze-out times
($\sim 8.3$ fm/c) and transverse radii ($\sim 4.5$ fm) for
central collisions.
The $\Lambda+\Sigma^0$ exhibits a similar impact parameter dependence
because of its associated production with kaons via the reaction
$(\pi,\rho,\omega)(N,\Delta,N^*)\to K(\Lambda,\Sigma)$. Except for the
very central collisions the $\Lambda+\Sigma^0$ yield in the ART model
is found to be consistent with the E895 data.

The $\Xi^-$ yield obtained by strangeness-exchange reactions between
antikaons and lambda/sigma for different impact parameters is also shown in
Fig. \ref{bmpact}. Similar to the singly strange hadrons, the $\Xi^-$
yield exhibits nearly a linear dependence on impact parameter.
This is due to the fact that $\Xi^-$s are mostly produced
from secondary collisions in the ART model. Fig. \ref{bmpact}
further shows that while the $\Lambda+\Sigma^0$ is enhanced by a factor of 4,
the $\Xi^-$ yield grows steadily by a factor of 6 from peripheral to
central collisions. This specific hierarchy of strangeness enhancement
of $\Xi > \Lambda+\Sigma^0$ has been reported for Pb+Pb collision at the SPS by
the WA97 collaboration \cite{ander}. However, at SPS energies, the $\Xi^-$
yield tends to saturate for a large number of participants 
$N_{\rm part} > 100$, and its enhancement (by $\sim 3$) in central 
collisions is smaller than our prediction for AGS energies. The 
steady growth of $\Xi^-$ with centrality is a clear indication that 
its yield is sensitive to the volume of the fireball formed in 
the collision. Furthermore, we find that the
$\Xi^-$ yield from strangeness-exchange reactions in the ART model is in
reasonable agreement with the E895 data \cite{paul}. This is in contrast
to SPS energies, where the cascade from the transport model was 
found \cite{pal1} to underpredict somewhat the WA97 data, especially 
for central collisions.

\subsection{Comparisons with the statistical model}

It may be interesting to find if strange hadrons also reach chemical 
equilibrium in heavy ion collisions at $6A$ GeV. To this end we
compare results from the transport model with those from the 
statistical model based on the grand canonical ensemble 
\cite{braun,cley} with complete thermal, chemical, and strangeness 
equilibrium. In this model, the particle density in the 
Boltzmann approximation is given by
\begin{eqnarray}\label{pdens}
\frac{N_i}{V}=\frac{gm_i^2T}{2\pi^2} K_2(m_i/T) e^{\mu_i/T},
\end{eqnarray}
in the usual notation \cite{braun,cley}.
Assuming $\Xi$ to be also in chemical equilibrium, the cascade to nucleon
ratio is
\begin{eqnarray}\label{cn}
&&\frac{N_\Xi}{N_N} = \left(\frac{m_\Xi}{m_N} \right)^2
\frac{K_2(m_\Xi/T)}{K_2(m_N/T)} e^{2\mu_s}  \nonumber \\
&=&4 \left(\frac{N_\Lambda}{N_N} \right)^2
\left(\frac{m_\Xi m_N}{m^2_\Lambda} \right)^2
\frac{K_2(m_\Xi/T) K_2(m_N/T)}{[K_2(m_\Lambda/T)]^2}.
\end{eqnarray}
For central Au+Au collisions, we use the ART model result for the
ratio $N_\Lambda/N_N = 0.025$. This gives a $\Xi^-$ yield of
$N_{\Xi^-} = 0.35-0.38$ at a freeze-out temperature of $T=120-170$ MeV. 
Note that the production of the strange particle $\Xi$ is less than 
1 per event. Thus, strangeness conservation must be implemented 
locally, i.e., a canonical instead of grand canonical ensemble must 
be used \cite{grand}. Therefore the present estimate of $\Xi^-$ gives 
only an upper bound. The higher value for the estimate of $\Xi^-$ 
from the statistical model compared to that obtained in the present 
transport approach indicates that $\Xi^-$ does not reach chemical 
equilibrium. Furthermore, like other strange hadrons, $\Xi$ freezes 
out relatively earlier than nonstrange hadrons with a mean
time and transverse radius of 8.4 fm/c and 4.6 fm when the central
energy density is about $\epsilon \approx 1$ GeV/fm$^3$.

The $\Xi^-/(\Lambda+\Sigma^0)$ ratio of 0.026 for $b=0-3$ fm in
the ART model is smaller than the equilibrium model estimate of
0.042 but slightly larger than the E895 data \cite{paul}. 
At midrapidity, the $\Xi^-/(\Lambda+\Sigma^0)$ ratio of 
0.12 has been observed by the E810 Collaboration
for central Si+Au collisions at $14.6A$ GeV \cite{eise}. Since
$\Xi$ is generated from $\Lambda,\Sigma$ by strangeness-exchange
reactions, its yield would naturally be more peaked at midrapidity
than that for $\Lambda$s. This is corroborated by a higher 
$\Xi^-/(\Lambda+\Sigma^0)$ ratio of 0.03 at midrapidity $|y|<0.50$.

Another way to see if the yield of $\Xi$ approaches that for
chemical equilibrium is to study how it changes when the cross section
is artificially increased for the reaction
$\bar K(\Lambda,\Sigma) \rightleftharpoons\pi \Xi$. This method was
introduced in Ref. \cite{hart} to study whether or not $K^-$ reaches
chemical equilibrium at energies available at SIS/GSI. The latter energies 
are below the threshold for $K^-$ production from nucleon-nucleon 
collisions in free space. Figure \ref{bmpact} shows that the $\Xi^-$ 
yield increases rather dramatically with increasing cross section 
$\sigma$. Equally important is the fact that for a given $\sigma$, 
the enhancement is even more pronounced for central collisions. 
Nonetheless, it is only for very large cross sections 
($\sim 6\sigma$), where the hydrodynamical limit may have been
reached, that the $\Xi^-$ approaches chemical equilibrium with 
$N_{\Xi^-} = 0.55$.

Interestingly, this value of $\Xi^-$ is even larger than that predicted
from the statistical model. Possible reasons for this result are: (i)
The temperature at complete chemical equilibrium, if reached at all,
is much higher than that in the statistical model; (ii) The increased
cross section has altered the reaction dynamics especially for $\bar K$
and $\Lambda$, i.e., a large abundance of $\bar K$s and $\Lambda+\Sigma^0$s 
so produced will increase the $\Xi$ at a higher temperature;  or (iii) 
large numbers of $\Xi$s, produced at an early
and dense stage do not have sufficient time to be annihilated by
rescattering because of rapid expansion of the system during its
subsequent evolution. The possibility of large $\Xi$ production from
case (i) alone is remote as its yield in the present statistical model
is found to be rather insensitive to temperature.

\subsection{Excitation functions}

\begin{figure}[ht]
\centerline{\epsfig{file=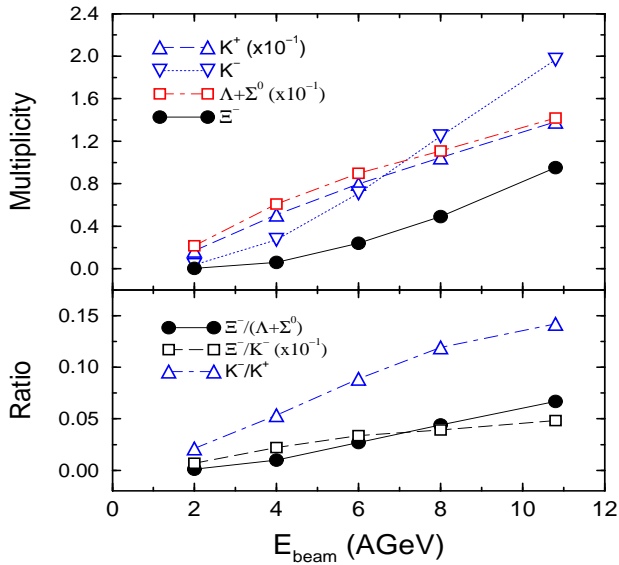,width=3.2in,height=3.0in,angle=0}}
\vspace{-0.1cm}
\caption{Excitation function for the total yields of $K^+$, $K^-$, 
$\Lambda+\Sigma^0$ and $\Xi^-$ (top panel) and the ratios
$K^-/K^+$, $\Xi^-/(\Lambda+\Sigma^0)$ and $\Xi^-/K^-$ (bottom panel) 
in central ($b=0-3$ fm) Au+Au collisions from ART model.}
\label{elab}
\end{figure}

In Fig. \ref{elab}, we show the excitation function for the strange particle
abundances (top panel) in central Au+Au collisions from the ART model.
As expected, with increasing beam energy the multiplicities of the hadrons
increases. The enhancement is more pronounced for antikaons and cascades
as the energy available for these massive and rare particle production
increases. This is clearly evident in the bottom panel of the Fig. \ref{elab}
where the $K^-/K^+$ ratio is found to increase monotonically with beam
energy. The $\Xi^-$ has the largest rate of increase with energy resulting 
in an increasing of the ratios $\Xi^-/(\Lambda+\Sigma^0)$ and $\Xi^-/K^-$ 
with beam energy.

\section{summary}

In summary, we have used a relativistic hadronic transport model to
investigate the doubly strange cascade production. We find that
the strangeness-exchange reactions between antikaons and hyperons lead
to substantial production of $\Xi^-$ in Au+Au collisions at $6A$ GeV
that is in reasonable agreement with the data. This is in stark contrast to
that at SPS energies, where the strangeness-exchange reactions underpredict
the data for the cascade yield in central collisions. The success of the 
purely hadronic scenario eliminates the scope of an exotica that may be 
necessary to explain the $\Xi$ yield at the AGS energies. Among all 
the strange hadrons, the $\Xi$ particle abundances reveal the strongest
enhancement with centrality of collision and also with increasing beam energy.


\begin{acknowledgements}
This paper was based on work supported in part by the US National 
Science Foundation under Grant No. PHY-0098805, the Welch Foundation 
under Grant No. A-1358 (C.M.K and S.P), and the U.S. Department of 
Energy under grant DE-FG02-87ER40331 (J.A, P.C, and R.L).
\end{acknowledgements}

\end{multicols}
\end{document}